\documentclass[preprintnumbers,article,amsmath,amssymb,floatfix,10pt,prd,twocolumn,
superscriptaddress,nofootinbib]{revtex4-2}
\usepackage{bm}
\usepackage{amsfonts}
\usepackage{latexsym}
\usepackage[latin1]{inputenc}
\usepackage{graphicx}
\usepackage{amsmath}
\usepackage{palatino}
\usepackage{mathpazo}
\usepackage{textcomp}
\usepackage{float}
\usepackage{booktabs}
\usepackage{dcolumn}
\usepackage{ragged2e}
\usepackage{hyperref}
\hypersetup{colorlinks,citecolor=blue}
\hypersetup{colorlinks=true,linkcolor=blue,filecolor=magenta,    urlcolor=blue}
\usepackage{amsmath}
\usepackage{subfigure}
\usepackage[]{natbib}
\usepackage{xcolor}
\usepackage{orcidlink}
\usepackage{epsfig}
\usepackage{caption}
\usepackage{subcaption}
\usepackage{commath}
\def\jnl@style{\it}
\def\aaref@jnl#1{{\jnl@style#1}}

\def\aaref@jnl#1{{\jnl@style#1}}

\def\aj{\aaref@jnl{AJ}}                   % Astronomical Journal
\def\apj{\aaref@jnl{ApJ}}                 % Astrophysical Journal
\def\apjl{\aaref@jnl{ApJ}}                % Astrophysical Journal, Letters
\def\apjs{\aaref@jnl{ApJS}}               % Astrophysical Journal, Supplement
\def\apss{\aaref@jnl{Ap\&SS}}             % Astrophysics and Space Science
\def\aap{\aaref@jnl{A\&A}}                % Astronomy and Astrophysics
\def\aapr{\aaref@jnl{A\&A~Rev.}}          % Astronomy and Astrophysics Reviews
\def\aaps{\aaref@jnl{A\&AS}}              % Astronomy and Astrophysics, Supplement
\def\mnras{\aaref@jnl{Mon.~Not.~Roy.~Astron.~Soc.}}             % Monthly Notices of the RAS
\def\prd{\aaref@jnl{Phys.~Rev.~D}}        % Physical Review D
\def\prc{\aaref@jnl{Phys.~Rev.~C}}  % Physical Review C
\def\prl{\aaref@jnl{Phys.~Rev.~Lett.}}    % Physical Review Letters
\def\qjras{\aaref@jnl{QJRAS}}             % Quarterly Journal of the RAS
\def\skytel{\aaref@jnl{S\&T}}             % Sky and Telescope
\def\ssr{\aaref@jnl{Space~Sci.~Rev.}}     % Space Science Reviews
\def\zap{\aaref@jnl{ZAp}}                 % Zeitschrift fuer Astrophysik
\def\nat{\aaref@jnl{Nature}}              % Nature
\def\aplett{\aaref@jnl{Astrophys.~Lett.}} % Astrophysics Letters
\def\apspr{\aaref@jnl{Astrophys.~Space~Phys.~Res.}} % Astrophysics Space Physics Research
\def\physrep{\aaref@jnl{Phys.~Rep.}}      % Physics Reports
\def\physscr{\aaref@jnl{Phys.~Scr}}       % Physica Scripta
\def\commat{\aaref@jnl{Comm.~Math.~Phys.}}              % Communications in Mathematical Physics
\def\science{\aaref@jnl{Science}}               % Science
\def\cqg{\aaref@jnl{Classical Quant.~Grav.}}            % Classical and Quantum Gravity
\def\jpcs{\aaref@jnl{JPCS}}                                     % Journal of Physics Conference Series
\def\ijmpd{\aaref@jnl{Int.~J.~Mod.~Phys.~D}}                    % International Journal of Modern Physics D
\def\grg{\aaref@jnl{Gen.~Relat.~Gravit.}}               % General Relativity and Gravitation
\def\rpp{\aaref@jnl{Rep.~Prog.~Phys.}}          % Reports on Progress in Physics
\def\npa{\aaref@jnl{Nucl.~Phys.~A}}        % Nuclear Physics A
\def\lrr{\aaref@jnl{Living Rev.~Rel.}}                   % Living reviews in relativity
\def\jcap{\aaref@jnl{J.~Cosmology Astropart.~Phys.}}    % Journal of cosmology and astroparticle physics
\def\rmp{\aaref@jnl{Rev.~Mod.~Phys.}}   %Reviews of modern physics
\def\epjc{\aaref@jnl{Eur.~Phys.~J.~C}} 
\def\plb{\aaref@jnl{~Phy.~Lett.~B}} 
\def\mpla{\aaref@jnl{Mod.~Phy.~Lett.~A}} 
\def\arxiv{\aaref@jnl{arxiv.org}}

\allowdisplaybreaks[1]
\begin{document}
%\color{red}
\color{black}       %% For one column
\title{Constraints on $\Lambda(t)$CDM cosmology using Cosmic Chronometers and Supernova data}

\author{Yerlan Myrzakulov\orcidlink{0000-0003-0160-0422}}\email[Email: ]{ymyrzakulov@gmail.com} 
\affiliation{Department of General \& Theoretical Physics, L.N. Gumilyov Eurasian National University, Astana, 010008, Kazakhstan.}

\author{M. Koussour\orcidlink{0000-0002-4188-0572}}
\email[Email: ]{pr.mouhssine@gmail.com}
\affiliation{Department of Physics, University of Hassan II Casablanca, Morocco.}

\author{M. Bulanbay}\email[Email: ]{bulanbai.merei@gmail.com} 
\affiliation{Department of General \& Theoretical Physics, L.N. Gumilyov Eurasian National University, Astana, 010008, Kazakhstan.}

\author{S. Muminov\orcidlink{0000-0003-2471-4836}}
\email[Email: ]{sokhibjan.muminov@gmail.com}
\affiliation{Mamun University, Bolkhovuz Street 2, Khiva 220900, Uzbekistan.}

\author{J. Rayimbaev\orcidlink{0000-0001-9293-1838}}
\email[Email: ]{javlon@astrin.uz}
\affiliation{Institute of Fundamental and Applied Research, National Research University TIIAME, Kori Niyoziy 39, Tashkent 100000, Uzbekistan.}
\affiliation{University of Tashkent for Applied Sciences, Str. Gavhar 1, Tashkent 100149, Uzbekistan.}
\affiliation{Urgench State University, Kh. Alimjan Str. 14, Urgench 221100, Uzbekistan}
\affiliation{Shahrisabz State Pedagogical Institute, Shahrisabz Str. 10, Shahrisabz 181301, Uzbekistan.}
%
%%%%%%%%%%%%%%%%%%%%%%%%%%%%%%%%%%%%%  DATE  %%%%%%%%%%%%%%%%%%%%%%%%%%%%%%%%%%%%
%\date{\today}
\begin{abstract}

In this manuscript, we investigate the constraints on dynamical vacuum models within the framework of $\Lambda(t)$CDM cosmology by assuming a parameterization of the vacuum energy density as $\rho_{\Lambda}(t)=\rho_{\Lambda 0} \left[1 + \alpha (1 - a)\right]$, where $\rho_{\Lambda 0}$ is the present vacuum density and $\alpha$ is a free parameter. We use 31 cosmic chronometer data points and 1048 Pantheon type Ia supernova samples to constrain the model parameters. Our statistical analysis employs Markov Chain Monte Carlo (MCMC) simulations. We have found that the universe is currently undergoing accelerated expansion, transitioning from a decelerating phase. The transition redshift $z_t=0.65^{+0.03}_{-0.19}$ obtained from the combined CC+SNe dataset is consistent with recent constraints. The total EoS indicates an accelerating phase, with density parameters for matter and vacuum energy exhibiting expected behaviors. The $Om(z)$ diagnostic shows distinct behaviors for different datasets, and the present value of the jerk parameter deviates slightly from the $\Lambda$CDM model but remains consistent within uncertainties. These findings support the dynamic nature of dark energy and provide valuable constraints on the evolution of the universe.

\end{abstract}

\maketitle

\tableofcontents

\section{Introduction}\label{sec1}

Observations accumulated over the years provide strong evidence supporting the current acceleration of the universe. This compelling evidence suggests the existence of a generic cause responsible for this acceleration, which is commonly referred to as dark energy (DE). Key observations supporting this conclusion include those by Riess et al. \cite{Riess/1998}, Perlmutter et al. \cite{Perlmutter/1999}, the WMAP collaboration \cite{WMAP/2003}, and the Planck collaboration in several papers \cite{Planck/2014,Planck/2015,Planck/2016,Planck/2020}, among others referenced therein. These findings have significantly shaped our understanding of cosmology, emphasizing the need to explore the nature and properties of DE to unravel the mysteries of the universe's expansion. Despite the dedicated efforts of cosmologists, the physical nature of DE remains elusive, leaving us ignorant of the ultimate cause of the observed acceleration of the universe. This fundamental mystery continues to challenge our understanding of fundamental physics and the nature of cosmic evolution. The quest to decipher the DE code remains at the forefront of cosmological research, driving scientists to explore novel theories, observational techniques, and experimental approaches in pursuit of a comprehensive understanding of this enigmatic cosmic phenomenon. 
One of the significant theoretical challenges in cosmology is the cosmological constant (CC) problem, which has been extensively discussed in the literature \cite{Weinberg/1989,Sahni/2000,Padmanabhan/2003,Peebles/2003,Copeland/2006}. The CC, denoted by $\Lambda$, or equivalently, the vacuum energy density associated with it, $\rho_{\Lambda} = \Lambda$, is often considered the simplest explanation for DE. However, its theoretical implications and the discrepancy between its predicted and observed values pose significant challenges to our understanding of fundamental physics and the nature of the universe's expansion. The CC problem remains a central issue in cosmology, driving ongoing research efforts to reconcile theoretical predictions with observational evidence.

Historically, the CC was first introduced by A. Einstein in the gravitational field equations a century ago \cite{Einstein/1917}. This positive, constant, and exceedingly small value, typically of the order of $\rho_{\Lambda} \sim 2.7\times 10^{-47}$ GeV$^4$ in particle physics units, has been proposed to explain the observed acceleration of our cosmos. This tiny value is consistent with observations and is a key component of the standard cosmological model, known as the $\Lambda$CDM model. In the $\Lambda$CDM model, the CC $\Lambda$ is assumed to remain constant throughout the history of the universe, with $\Omega_\Lambda \simeq 0.7$ and $\Omega_m \simeq 0.3$ at present \cite{Planck/2020}. This model, which also includes the hypothesis of dark matter (DM), has been remarkably successful in explaining a wide range of cosmological observations, from cosmic microwave background radiation to large-scale structure formation. However, the physical origin of the CC and its extremely small but nonzero value remain among the most profound puzzles in modern cosmology. There is a growing consensus that to gain a deeper understanding of the nature of DE and DM, it is essential to explore more complex models. For example, scenarios that incorporate interactions within the dark sector are being increasingly considered \cite{Amendola/2000,Maia/2002,Koivisto/2005,Lee/2006,Bertolami/2007}.

In this context, the simplest examples of interacting DE/DM models are scenarios involving vacuum decay ($\Lambda(t)$CDM). In these cosmologies, the DE pressure-to-energy density ratio, $\omega_{\Lambda}$, is exactly $-1$ \cite{Ozer/1986,Freese/1987,Chen/1990,Berman/1991,Pavon/1991,Carvalho/1992,Arbab/1994,Barrow/2006,Wang/2006,Montenegro/2007,Overduin,Rezaei}. This type of model is based on the concept that DE arises from vacuum quantum fluctuations in curved spacetime, following a renormalization process that subtracts the divergent vacuum contribution in flat spacetime. The resulting effective vacuum energy density depends on spacetime curvature, decreasing from high initial values to smaller ones as the universe expands \cite{Carneiro/2003}. Due to the conservation of total energy, as implied by Bianchi identities, the variation in vacuum density leads either to particle production or an increase in the mass of DM particles, which are general features of decaying vacuum or, more broadly, interacting DE models \cite{Alcaniz/2005}. These models offer intriguing insights into the dynamic interplay between DM and DE, shedding light on the evolution of the universe and the fundamental nature of these cosmic components.

In this paper, we delve deeper into our investigation and explore new observational implications of the $\Lambda(t)$CDM scenario outlined earlier. To achieve this, we use distance measurements from the cosmic chronometer sample and data from the Type Ia supernovae sample. Our analysis reveals that, in addition to the intriguing cosmic history associated with this class of $\Lambda(t)$CDM models, the current observational data only marginally favors a conventional, spatially flat $\Lambda$CDM model over them. This suggests that the $\Lambda(t)$CDM framework offers a compelling alternative that warrants further exploration and scrutiny in light of ongoing advancements in observational techniques and theoretical developments. As mentioned earlier, our study is motivated by the possibility that the inherent dynamics in $\Lambda(t)$ are linked to fundamental principles of quantum field theory (QFT). Among the dynamical vacuum models, the running vacuum model (RVM) is a prominent model under investigation, as it offers connections to significant aspects of QFT in curved spacetime. This framework, as described in \cite{Sola/2008,Sola/2011,Sola/2013} and references therein, replaces the $\Lambda$CDM model with the $\Lambda(t)$CDM model, where $\Lambda(t)=\Lambda(H)$, or equivalently $\rho_\Lambda(t)=\rho_\Lambda(H)$, plays the role of running vacuum energy density. S. Carneiro et al. \cite{Carneiro/2008} discussed the cosmological implications of a scenario characterized by a varying cosmological term, where the vacuum energy density diminishes linearly with the Hubble parameter ($\Lambda(t)=\sigma H$). The structure of our work is outlined as follows: In Sec. \ref{sec2}, we provide a detailed description of the dynamic model of vacuum energy. Sec. \ref{sec3} is dedicated to presenting the observational constraints on the model parameters of vacuum energy. In Sec. \ref{sec4}, we delve into a discussion of the physical properties of the model. Lastly, in Sec. \ref{sec5}, we present our conclusions based on the findings and analyses presented in the preceding sections.

\section{Dynamic model of vacuum energy}\label{sec2}

In this section, we outline the fundamental equations of a general cosmological scenario. Throughout our analysis, we focus on the gravitational field equations within the framework of GR, expressed as
\begin{equation}
    G_{\mu \nu}=\kappa \Tilde{ T}_{\mu \nu },
\end{equation}
where $\kappa=8 \pi G=1$, $G_{\mu \nu}=R_{\mu \nu}- \frac{1}{2} R g_{\mu \nu}$ is the Einstein tensor and $\Tilde{T}_{\mu\nu}\equiv T_{\mu\nu}+g_{\mu\nu} \rho_{\Lambda}$ represents the total energy-momentum tensor, which accounts for the contributions of both matter and a time-dependent vacuum term, with energy density $\rho_\Lambda=\Lambda(t)$. In this paper, we adopt the assumption that the vacuum behaves like a perfect fluid with an EoS given by $\rho_\Lambda=-p_\Lambda$. When matter can also be modeled as a perfect fluid and is distributed uniformly and isotropically, as suggested by the cosmological principle, we can express
\begin{equation}
    \Tilde{T}_{\mu\nu}= (\rho_{\Lambda}-p_{m})\,g_{\mu\nu}+(\rho_{m}+p_{m})u_{\mu}u_{\nu},
\end{equation}
where $u_{\mu}$ represents the $4$-velocity of the cosmic fluid, $\rho_m$ is the energy density of matter, and $p_m$ is isotropic pressure.

In addition, we adopt the standard cosmological framework, which is based on the Friedmann-Lema\^{i}tre-Robertson-Walker (FLRW) metric \cite{Ryden},
\begin{equation} \label{FLRW}
ds^2 = dt^2 - a^2(t) \left[ dr^2 + r^2(d\theta^2 + \sin^2\theta d\phi^2) \right], 
\end{equation}
where $t$ is the cosmic time and $a(t)$ is the scale factor of the universe. However, we assume the possibility of matter interacting with the vacuum. This interaction, as described by the continuity equation, yields
\begin{equation}
 \dot{\rho}_{m} + 3H(\rho_{m}+p_{m}) = -\dot{\Lambda}(t), 
\end{equation}
which suggests that the conservation of matter is not independent, as the decaying vacuum serves as a source of matter. In this scenario, the vacuum's decay contributes to the production or transformation of matter, altering the traditional conservation laws and leading to a more intricate understanding of the dynamics between matter and the vacuum.

Through the application of the metric and the gravitational field equations, we find that the standard Friedmann and acceleration equations for the current universe maintain a formal identity with those of the standard $\Lambda$CDM paradigm \cite{Koussour1,Myrzakulov1},
\begin{eqnarray}
\label{F1}
3H^{2}&=&\rho_{m}+\rho_{\Lambda}, \\
2{\dot{H}}+3H^{2}&=&-p_{\Lambda}, 
   \label{F2} 
\end{eqnarray}
where $H=\frac{\dot{a}}{a}$ is the Hubble parameter, which represents the current rate of expansion of the universe.

From the above equations, we can express the evolution equation as 
\begin{equation} \label{F22}
2{\dot{H}}+3H^{2}-\Lambda(t)=0.
\end{equation}

The interesting work by Wang and Meng \cite{Wang/2017} motivates us to adopt the following parameterization for the vacuum energy density:
\begin{equation}
   \rho_{\Lambda}(t)=\rho_{\Lambda 0} \left[1 + \alpha (1 - a)\right],
\end{equation}
where $\rho_{\Lambda 0}$ represents the present vacuum density, while $\alpha$ serves as a free parameter in the vacuum model. It is noteworthy that this model simplifies to the $\Lambda$CDM case $\rho_{\Lambda}(t)=\rho_{\Lambda 0}=\Lambda$ when $\alpha=0$. Determining the precise values of $\alpha$ will involve comparing the model's predictions with the most recent cosmological observations. For this purpose, rather than using the conventional time variable $t$, we introduce the redshift $z$ as the independent variable, defined as 
\begin{equation}
    1+z=\frac{a_0}{a(t)},
\end{equation}
where $a_0 = 1$ denotes the present value of the scale factor. Therefore, we can substitute derivatives with respect to time with derivatives with respect to redshift using the following relation:
\begin{equation}
\label{dt}
    \frac{d}{dt}=-H(z)(1+z) \frac{d}{dz}.
\end{equation}

Thus, the time derivative of the Hubble parameter can be represented as
\begin{equation}
\label{dt}
\dot H=-H(z)(1+z) \frac{dH(z)}{dz}.
\end{equation}

Now, we define the dimensionless density parameters for matter and vacuum energy as
\begin{equation}\label{3j}
\Omega_{m}=\frac{\rho_{m}}{\rho_c}, \ \  \   \Omega_{\Lambda}=\frac{\rho_{\Lambda}}{\rho_c},
\end{equation}
where $\rho_c=3 H^2$.

Therefore, the relation between dimensionless density parameters can be derived from Eq. (\ref{F1}) as follows:
\begin{equation}
\Omega_{m}+\Omega_{\Lambda}=1.
\end{equation}

Under the conditions $H>0$ and $\rho_{m}>0$, integrating Eq. (\ref{F22}) yields the following expression for the Hubble parameter in terms of redshift $z$:
\begin{equation}
\begin{split}
H(z) = H_{0}\Bigg[ & \Bigg( 1-\frac{\left( 1-\Omega _{m0}\right) \left( 4+\alpha \right) }{4} \Bigg) \\
& \times \left( 1+z\right) ^{3} + \frac{\left( 1-\Omega _{m0}\right) \left( 4+\alpha +4z\left( 1+\alpha \right) \right) }{4\left( 1+z\right) } \Bigg] ^{\frac{1}{2}},
\label{Hz}
\end{split}
\end{equation}
where $H_0=H(z=0)$ denotes the current value of the Hubble parameter, while $\Omega_{m0}$ represents the current value of the matter density parameter. 

From Eq. (\ref{Hz}), it is clear that if $\Lambda=0$ and $\Omega_{m0}=1$, the above expression reduces to the Einstein-de Sitter solution, which is represented by $H(z)=H_0 (1+z)^{3/2}$ \cite{Carneiro/2008}. This solution describes a matter-dominated universe with no cosmological constant, where the expansion rate is solely determined by the matter content. On the other hand, if $\Lambda \neq 0$ and $\Omega_{m0} = 0$, the expression tends toward the de Sitter universe, where \( H(z) = \sqrt{\frac{\Lambda}{3}} \). In this scenario, the universe is dominated by the cosmological constant, leading to an accelerated expansion rate with time. This model is characterized by a universe that approaches a state of exponential expansion, driven primarily by the repulsive effect of the cosmological constant.

Please take note that, as a result of the matter production, the expression in Eq. (\ref{Hz}) differs significantly from that of the standard $\Lambda$CDM case, which encompasses both a cosmological constant and matter. Specifically, when $\alpha=0$, the expression for $H(z)$ simplifies to the $\Lambda$CDM model,
\begin{equation}
    H(z)=H_0 \sqrt{\Omega_{m0}(1+z)^3+(1-\Omega_{m0})}.
\end{equation}

Therefore, the estimated value of $\alpha$ will indicate whether cosmological observations favor a cosmological constant in the case of a non-decaying or decaying vacuum.

In the following section, we will endeavor to derive the values of the model parameters by analyzing the most recent cosmological dataset available. 

\section{Observational constraints on model parameters of vacuum energy} \label{sec3}

In this section, we delve into the methodology employed and the diverse observational samples utilized to constrain the parameters of the cosmological model under consideration ($H_0$, $\Omega_{m0}$, $\alpha$). Specifically, we employ a Markov Chain Monte Carlo (MCMC) method for statistical analysis and to derive the posterior distributions of the mentioned parameters. The data analysis is conducted using the \textit{emcee} package in Python, which facilitates the implementation of the MCMC method for parameter estimation \cite{Mackey/2013}. The optimization of the best fits parameters is achieved through the maximization of the probability function given by
\begin{eqnarray}
    \mathcal{L}\propto \exp(-\chi^2/2),
\end{eqnarray}
where $\chi^2$ represents the chi-squared function (Bayesian chi-squared) \cite{baye}. Further details regarding the $\chi^2$ function for different data samples are presented in the subsequent subsections.

\subsection{Sample of Cosmic Chronometers (CC)}

The Cosmic Chronometers (CC) method employs the differential aging technique to study the oldest and least active galaxies that are closely separated in redshift, in order to determine the Hubble rate. The method is based on the definition of the Hubble rate $ H = -\frac{1}{1+z}\frac{dz}{dt}$ for an FLRW metric. The CC method is a valuable tool for cosmological model analysis because it can calculate the Hubble parameter without relying on specific cosmological assumptions. Here, we use 31 CC data points within the redshift range $(0.1, 2)$ obtained from various surveys \cite{Jimenez:2003iv,Simon:2004tf,Stern:2009ep,Moresco:2012jh,Zhang:2012mp,Moresco:2015cya,Moresco:2016mzx,Ratsimbazafy:2017vga}. For the CC data, the $\chi^2$ function is defined as
\begin{equation}
    \chi_{CC}^2 = \Delta H^T (C_{CC}^{-1})\Delta H,
\end{equation}
where $C_{CC}$ is the covariance matrix components representing errors related to the observed $H(z)$ values, and $\Delta H$ is the vector representing the difference between predicted and measured values of $H(z)$ for each redshift data. The Hubble parameter's theoretical value can be calculated with the use of Eq. \eqref{Hz}.

\subsection{Sample of Type Ia Supernovae (SNe)}

SNe samples play a crucial role in cosmological studies due to their ability to provide information about the underlying geometry and dynamics of the Universe. By observing the brightness and redshift of Type Ia supernovae, researchers can discern valuable information about the expansion history of the cosmos, leading to significant advancements in our understanding of fundamental cosmological parameters and the nature of DE. For this analysis, we use the extensive Pantheon dataset, which consists of 1048 data points spanning the redshift range $0.01 < z < 2.26$ \cite{pantheon}. This comprehensive compilation integrates observations from multiple surveys, including SNLS, SDSS, HST, and Pan-STARRS1. The $\chi^2$ function corresponding to the SNe simple is defined as
\begin{equation}\label{4b}
\chi^2_{SNe}=\sum_{i,j=1}^{1048}\Delta\mu_{i}\left(C^{-1}_{SNe}\right)_{ij}\Delta \mu_{j},
\end{equation}
where $C_{SNe}$ represents the covariance matrix \cite{pantheon}, and
\begin{align}\label{4c}
\quad \Delta\mu_{i}=\mu^{th}(z_i,\theta_s)-\mu_i^{obs}.
\end{align}
represents the disparity between the measured distance modulus, obtained from cosmic observations, and its theoretical counterpart calculated from the model using the given parameter space $\theta_s$. Here, $\mu_i^{th}$ and $\mu_i^{obs}$ denote the theoretical and measured distance modulus, respectively. In addition, the theoretical distance modulus $\mu_i^{th}$ is expressed as
\begin{equation}
\mu_i^{th}(z)=m-M=5\log d_l(z)+25.
\end{equation}

Here, $m$ and $M$ represent the apparent and absolute magnitudes of a standard candle, respectively. Further, the luminosity distance $d_l(z)$ is expressed as
\begin{equation}
d_L(z)=\frac{c(1+z)}{H_0}S_K\bigg(H_0\int^z_0\frac{dz^\ast}{H(z^\ast)}\bigg),
\end{equation}
where
\begin{equation}
S_K(x)=    \begin{cases}
      \sinh(x\sqrt{\Omega_K})/\Omega_K,\quad ~ \Omega_K >0\\
      x,\quad\quad\quad\quad\quad\quad\quad \quad ~~\Omega_K=0\\
      \sin (x \sqrt{|\Omega_K|})/|\Omega_K|,~~ \Omega_K<0
    \end{cases}\,.
\end{equation}

\subsection{Sample of CC+SNe}

To simultaneously analyze both the CC and SNe samples, we use the following likelihood and chi-square functions:
\begin{eqnarray}
\mathcal{L}_{CC+SNe} &=& \mathcal{L}_{CC} \times \mathcal{L}_{SNe},\\
\chi^{2}_{CC+SNe} &=& \chi^{2}_{CC} + \chi^{2}_{SNe}.
\end{eqnarray}

\begin{figure}[h]
    \centering
    \includegraphics[scale=0.4]{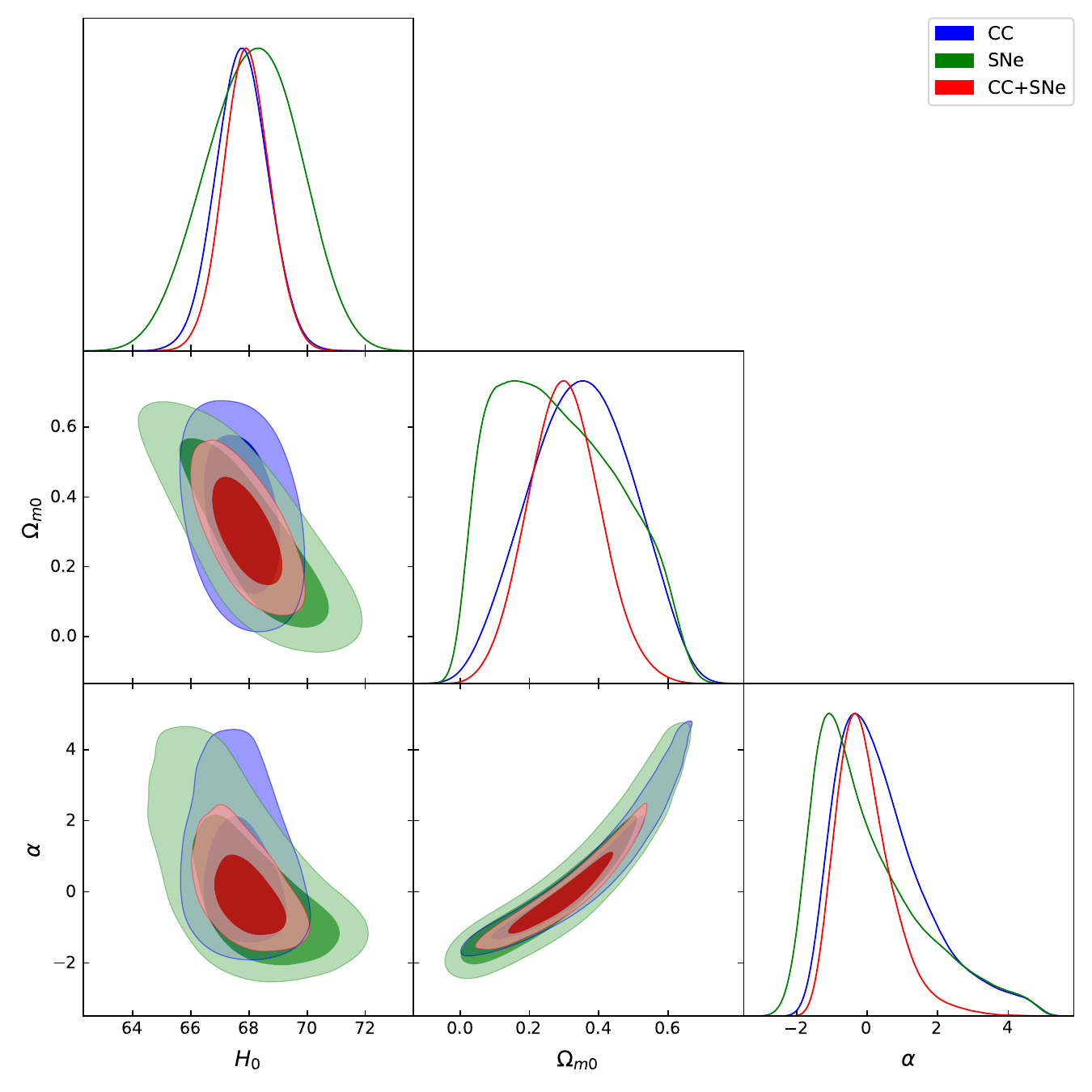}
    \caption{The regions corresponding to the $1-\sigma$ and $2-\sigma$ CL for the parameters are depicted based on the CC, SNe, and CC+SNe data.}
    \label{Con}
\end{figure}

For running the MCMC code, we employ 100 walkers and 1000 steps to determine the fitting results. The $1-\sigma$ and $2-\sigma$ confidence level (CL) contour plots are depicted in Fig. \ref{Con}, and the corresponding numerical results for the CC, SNe, and combined CC+SNe samples are presented in Tab. \ref{table}. In summary, our estimates for the Hubble constant are $H_0=67.8_{-1.7}^{+1.8}$ km/s/Mpc for CC data, $H_0=68.1^{+3.0}_{-3.1}$ km/s/Mpc for SNe data, and $67.92 \pm 0.80$ km/s/Mpc for the combined CC+SNe data. The $H_0$ tension refers to the discrepancies in the measurement of the Hubble parameter between the Planck collaboration and independent cosmological probes. The Planck collaboration \cite{Planck/2020} estimated $H_0 = 67.27 \pm 0.06$ km/sec/Mpc, while R19 \cite{Riess/2019} reported $H_0 = 74.03 \pm 1.42$ km/sec/Mpc. This difference leads to a tension of $4.4 \sigma$, indicating a significant discrepancy beyond what would be expected due to random variations or uncertainties alone. A detailed discussion of the $H_0$ tension and possible solutions can be found in Ref. \cite{Valentino/2021A}. When comparing our results with those of the Planck collaboration \cite{Planck/2020}, we observe that our estimated value of $H_0$ from joint CC+SNe data shows a small tension with the Planck results. Further investigations and analyses related to the $H_0$ tension are discussed in Refs. \cite{Yang/2021,Valentino/2021B}. 
In addition, our analysis yields accurate estimates for the matter density parameter, as detailed in Tab. \ref{table}. 

In this paper, we have further constrained $\alpha$ (which indicates a deviation from the $\Lambda$CDM model) to be within the range of $\alpha=0.4_{-2.1}^{+2.9}$, $\alpha=0.0_{-2.3}^{+3.4}$, and $\alpha=-0.04^{+0.53}_{-0.96}$ by analyzing the derived model with CC, SNe, and the combined compilation of CC+SNe data. It is noteworthy that $\alpha = 0$ for the SNe data, which corresponds to the $\Lambda$CDM model.

\begin{table}[h]
\begin{center}
\begin{tabular}{l c c c c}
\hline 
$dataset$              & $CC$ & $SNe$ & $CC+SNe$ \\
\hline
$H_{0}$ ($km/s/Mpc$) & $67.8_{-1.7}^{+1.8}$  & $68.1^{+3.0}_{-3.1}$  & $67.92 \pm 0.80$ \\

$\Omega_{m0}$   & $0.35_{-0.28}^{+0.27}$  & $0.28_{-0.27}^{+0.31}$  & $0.30 \pm 0.10$ \\

$\alpha$   & $0.4_{-2.1}^{+2.9}$  & $0.0_{-2.3}^{+3.4}$  & $-0.04^{+0.53}_{-0.96}$ \\

$q_{0}$   & $-0.48_{-0.42}^{+0.41}$  & $-0.58_{-0.42}^{+0.46}$  & $-0.55 \pm 0.15$ \\

$z_{tr}$   & $0.73_{-0.32}^{+0.03}$  & $0.73_{-0.06}^{+0.32}$  & $0.65^{+0.03}_{-0.19}$ \\

$\omega_{0}$   & $-0.65_{-0.28}^{+0.27}$  & $-0.72_{-0.27}^{+0.31}$  & $-0.70^{+0.10}_{-0.10}$ \\

$j_{0}$   & $0.61_{-0.22}^{+0.27}$  & $1.0_{-0.45}^{+0.09}$  & $1.04^{+0.48}_{-0.87}$ \\

\hline
\end{tabular}
%}
\caption{Numerical outcomes from the statistical analysis.}
\label{table}
\end{center}
\end{table}

\section{Physical properties of the model} \label{sec4}

\subsection{Deceleration parameter}

The deceleration parameter is defined as \cite{Koussour2,Koussour3,Koussour4,Koussour5},
\begin{equation}
    q =-1-\frac{\dot{H}}{H^2}.
\end{equation}

\begin{figure}[h]
    \centering
    \includegraphics[scale=0.7]{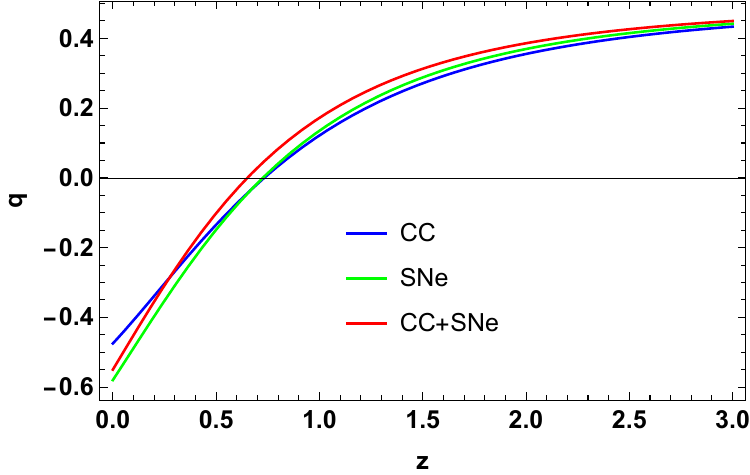}
    \caption{Profile of the deceleration parameter $q$ vs. cosmic redshift $z$.}
    \label{F_q}
\end{figure}

The positive value of the deceleration parameter signifies that the universe's expansion is slowing down, suggesting a decelerating phase. Conversely, a negative value indicates that the universe's expansion is accelerating, which aligns with the observed acceleration in the expansion of the universe. 
The plot depicting the variation of the deceleration parameter $q$ with respect to redshift $z$ is presented in Fig. \ref{F_q}, For the combined CC+SNe dataset, the transition redshift at which the universe transitions from a decelerating phase to an accelerating phase is determined to be $z_t=0.65^{+0.03}_{-0.19}$. Notably, this value is close to other recent constraints reported as $z_t=0.74 \pm 0.05$, $z_t=0.69^{+0.23}_{-0.12}$, and $z_t=0.60 ^{+0.21}_{-0.12}$ in the literature \cite{Farooq/2013,Lu/2011,Yang/2020}, respectively. Our analysis reveals that the universe in the proposed model is currently expanding in an accelerated phase, whereas in the earlier stages, it was in a decelerating expansion phase. Hence, the proposed model represents a transitioning universe, transitioning from a deceleration phase to the current phase of acceleration.

In this study, we constrain the present value of the deceleration parameter to be $q_0=-0.55 \pm 0.15$ (CC+SNe), which closely aligns with the value of $q_0$ obtained by \cite{Capozziello/2014,Capozziello/2020}. Fig. \ref{F_q} illustrates the dynamics of the deceleration parameter $q$ with respect to the redshift $z$ across the redshift range $z \in [0, 3]$. Furthermore, we have included the $1-\sigma$ CL values of the transition redshift and deceleration parameter for all datasets in Tab. \ref{table}.

\subsection{Total EoS and density parameters}

In this study, we have initially assumed that DE corresponds to vacuum energy with EoS $\omega_{\Lambda}=-1$. Now, we will proceed to investigate the total EoS of the model, which encompasses both matter and vacuum energy varying with time. This equation is expressed as
\begin{equation}
\omega=\frac{p_{\Lambda}}{\rho_{m}+\rho_{\Lambda}}=\frac{\omega_{\Lambda}}{1+\frac{\rho_m}{\rho_{\Lambda}}}.
    \label{EoS}
\end{equation}

\begin{figure}[h]
    \centering
    \includegraphics[scale=0.7]{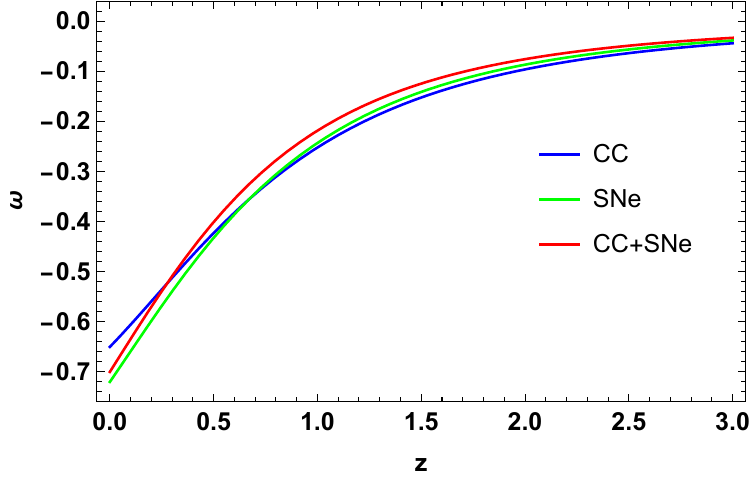}
    \caption{Profile of the total EoS parameter $\omega$ vs. cosmic redshift $z$.}
    \label{F_EoS}
\end{figure}

In the literature, the EoS parameter is crucial for distinguishing between the different epochs of accelerated and decelerated expansion in the universe. These epochs can be categorized as follows: (a) $\omega = 1$ represents a stiff fluid, which typically does not occur in the universe; 
(b) $\omega = \frac{1}{3}$ depicts the radiation-dominated phase, where radiation is the dominant component; 
(c) $\omega = 0$ indicates the matter-dominated phase, during which matter is the dominant component; and 
(d) $\omega < -\frac{1}{3}$ represents a decelerated expansion phase, characterized by negative pressure, possibly attributed to exotic forms of energy such as DE. The variation of the total EoS parameter is depicted in Fig. \ref{F_EoS}. It is evident from the figure that $\omega_0 < -\frac{1}{3}$, indicating an accelerating phase. The current values of the total EoS parameter are presented in Tab. \ref{table}, which are consistent with values reported in the literature \cite{Gruber/2014}.

\begin{figure}[h]
    \centering
    \includegraphics[scale=0.7]{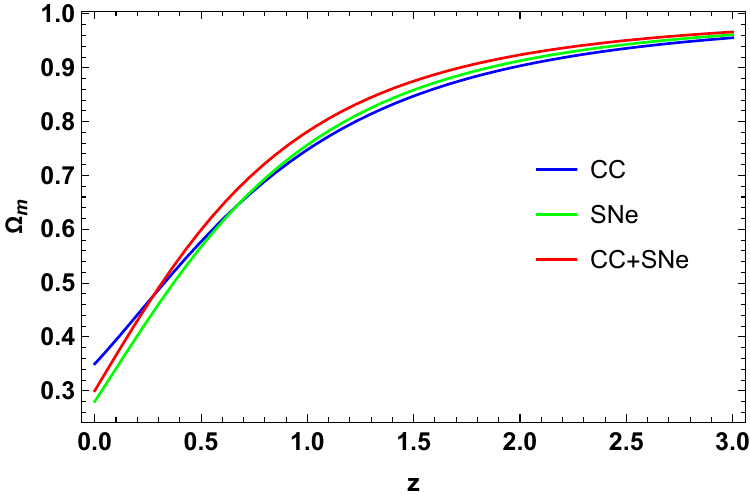}
    \caption{Profile of the density parameter for matter $\Omega_{m}$ vs. cosmic redshift $z$.}
    \label{F_Omega_m}
\end{figure}

\begin{figure}[h]
    \centering
    \includegraphics[scale=0.7]{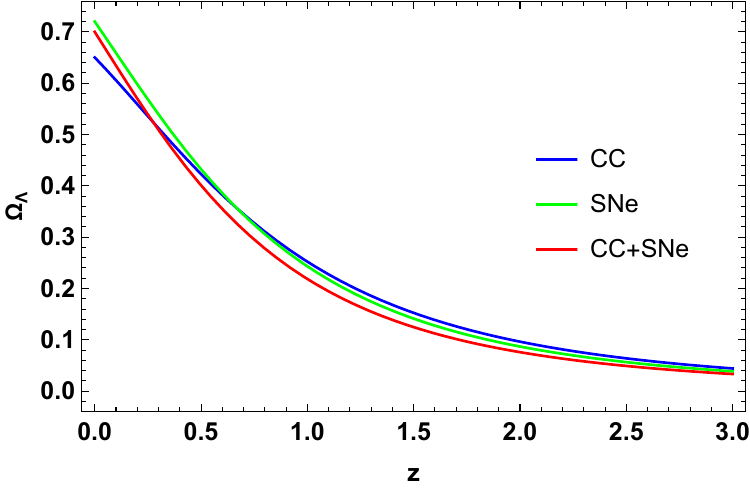}
    \caption{Profile of the density parameter for vacuum energy $\Omega_{\Lambda}$ vs. cosmic redshift $z$.}
    \label{F_Omega_L}
\end{figure}

Now, let's examine the behavior of the dimensionless density parameters for matter and vacuum energy, as expressed in Eq. (\ref{3j}), and illustrated in Figs. \ref{F_Omega_m} and \ref{F_Omega_L}. The present value of the density parameter for matter is determined to be $\Omega_{m0}=0.30 \pm 0.10$, which aligns with the results obtained from Planck observations \cite{Planck/2014,Planck/2015,Planck/2016,Planck/2020}, using the combined CC+SNe. The two graphs illustrate the well-established thermal evolution of the universe, which is marked by distinct epochs dominated first by matter and later by vacuum energy. This familiar pattern in the graphs reaffirms our understanding of the universe's evolution and is consistent with the standard cosmological model.

\subsection{$Om(z)$ diagnostic}

The $Om(z)$ diagnostic is defined as \cite{Sahni/2008,Sahni/2014},
\begin{equation}
Om\left( z\right) =\frac{\left( \frac{H\left( z\right) }{H_{0}}\right) ^{2}-1%
}{\left( 1+z\right) ^{3}-1}.    
\end{equation}

Here, $H_0$ represents the present value of the Hubble parameter. The negative slope of $Om(z)$ corresponds to quintessence-type behavior in the evolution of the DE component, where the DE density decreases with time. This behavior is associated with a scalar field that rolls down its potential slowly. On the other hand, a positive slope of $Om(z)$ corresponds to phantom behavior, where the DE density increases with time. This behavior is attributed to a scalar field with negative kinetic energy or a super-negative potential. The constant nature of $Om(z)$ represents the $\Lambda$CDM. In this model, the DE density remains constant over time, leading to a constant $Om(z)$ curve. In Ref. \cite{Shahalam/2015}, the authors investigated the $Om(z)$ diagnostic and found that its present value is zero for the $\Lambda$CDM universe.

\begin{figure}[h]
    \centering
    \includegraphics[scale=0.7]{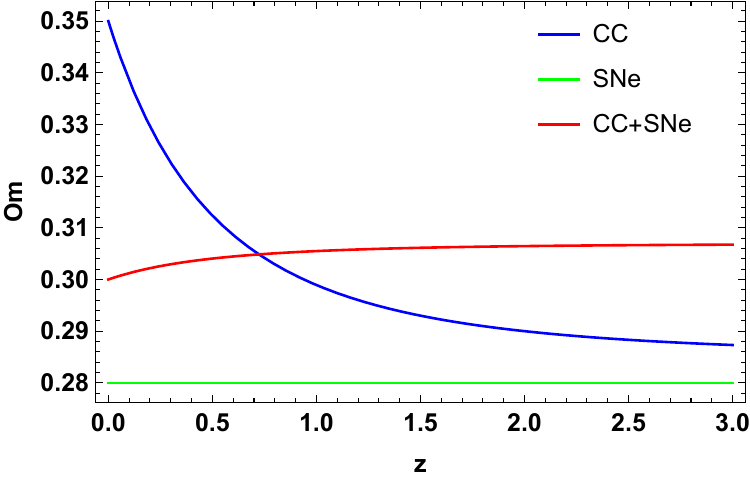}
    \caption{Profile of the $Om(z)$ diagnostic vs. cosmic redshift $z$.}
    \label{F_Om}
\end{figure}

The plot of $Om(z)$ with respect to redshift $z$ is depicted in Fig. \ref{F_Om}. Analysis of this plot reveals distinct behaviors for different datasets. Firstly, for the CC dataset, $Om(z)$ exhibits a decreasing trend with increasing redshift $z$, indicating a negative slope. This behavior corresponds to quintessence-type behavior. Secondly, when considering the SNe dataset, $Om(z)$ shows a constant slope throughout the evolution of the universe, which is characteristic of the $\Lambda$CDM model. Lastly, for the combined dataset, $Om(z)$ demonstrates an increasing trend with increasing redshift $z$, signifying a positive slope. This behavior aligns with phantom-type behavior.

\subsection{Jerk parameter}
The jerk parameter is defined as \cite{Sahni/2003,Alam/2003},
\begin{equation}
\label{jerk}
    j=\frac{\overset{...}{a}}{aH^{3}}=q(2q+1)+(1+z)\frac{dq}{dz}.
\end{equation}

\begin{figure}[h]
    \centering
    \includegraphics[scale=0.7]{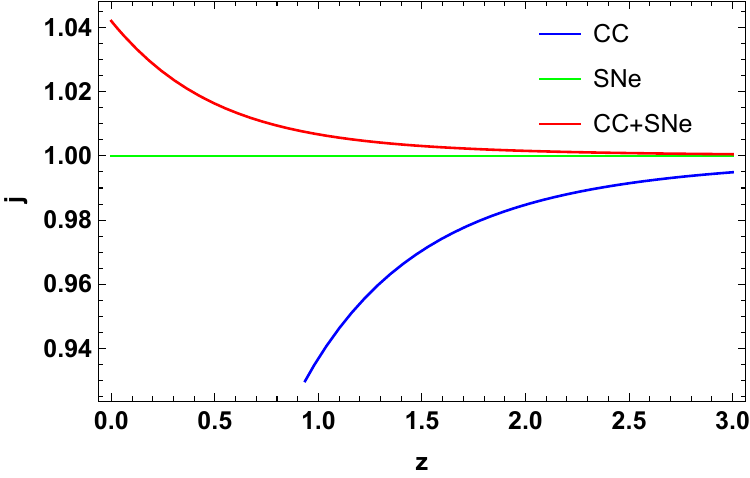}
    \caption{Profile of the jerk parameter $j$ vs. cosmic redshift $z$.}
    \label{F_j}
\end{figure}

In the $\Lambda$CDM model, the jerk parameter is expected to be constant over time, with its present value as $j_0 = 1$. The jerk parameter is frequently employed to distinguish between different DE models, particularly those with values deviating from $j_0 = 1$. Fig. \ref{F_j} shows the plot of the jerk parameter against redshift. In our study, we constrain the present value of the jerk parameter to be $j_0=1.04^{+0.48}_{-0.87}$ based on the combined CC+SNe dataset. It is important to note that, as mentioned earlier, in the case of the SNe dataset, the present value of $j_0$ corresponds to $\lambda$CDM model behavior. Mukherjee and Banerjee \cite{Mukherjee/2016} provided a parametric reconstruction of the jerk parameter using diverse observational datasets and constrained the model parameters through maximum likelihood analysis. On the other hand, Zhai et al. \cite{Zhai/2013} presented an extensive reconstruction of the jerk parameter. Both reconstructions accommodate the $\Lambda$CDM model well within the $1-\sigma$ CL. In Ref. \cite{Akarsu/2014}, Akarsu et al. introduced the hybrid expansion law and determined the present value of the jerk parameter to be $j_0=0.520 \pm 0.156$. In addition, other constraints on the jerk parameter can be found in Refs. \cite{Sahni/2003,Alam/2003,Mamon/2018}. Thus, given the current model (where $q_0 < 0$, $\omega_0<-\frac{1}{3}$ and $j_0>0$), it is clear that the dynamic DE model being studied is the most likely explanation for the universe's present acceleration, highlighting the importance of further research and scrutiny. 

\section{Conclusion} \label{sec5}

While it is evident that our universe is currently undergoing accelerated expansion, the precise nature of this phenomenon, known as CC ($\Lambda$) physics, remains largely elusive. The increasing availability of high-precision astronomical data underscores the urgency and complexity of unraveling the underlying physics governing CC. An essential query within this realm pertains to whether CC undergoes any temporal evolution. This pivotal question is central to our pursuit of understanding the fundamental characteristics of CC, which continues to captivate and challenge modern cosmology. Addressing this inquiry demands innovative theoretical paradigms and advanced observational methods capable of discerning the nuanced dynamics of CC throughout cosmic epochs.

In this study, we have investigated the dynamics of a specific category of $\Lambda(t)$CDM scenarios, characterized by the vacuum energy density $\rho_{\Lambda}(t)=\rho_{\Lambda 0} \left[1 + \alpha (1 - a)\right]$. We have also explored a potential methodology to differentiate these scenarios from the standard $\Lambda(t)$CDM model, focusing on their distinct features regarding the overall predicted cosmic evolution. Further, we have explored the dynamics of various cosmological parameters and diagnostics within the framework of a $\Lambda(t)$CDM cosmological model. Our analysis provides several key insights:
\begin{itemize}
    \item \textbf{Deceleration parameter:} We find that the universe is currently in an accelerated expansion phase, transitioning from a decelerating phase in earlier stages. The transition redshift $z_t=0.65^{+0.03}_{-0.19}$ obtained from the combined CC+SNe dataset is consistent with other recent constraints reported in the literature.
    \item \textbf{Total EoS and density parameters:} The total EoS parameter $\omega$ indicates an accelerating phase, consistent with our understanding of DE behavior. The density parameters for matter and vacuum energy exhibit expected behaviors, with $\Omega_{m0}=0.30 \pm 0.10$ aligning with Planck observations.
    \item \textbf{$Om(z)$ diagnostic:} The $Om(z)$ diagnostic reveals distinct behaviors for different datasets, with the CC dataset showing quintessence-like behavior, the SNe dataset consistent with $\Lambda$CDM, and the combined dataset indicating phantom-like behavior.
    \item \textbf{Jerk parameter:} The present value of the jerk parameter $j_0=1.04^{+0.48}_{-0.87}$ from the combined CC+SNe dataset deviates slightly from the $\Lambda$CDM model but remains consistent within uncertainties.
\end{itemize}
Overall, our analysis supports the dynamic nature of CC and provides valuable constraints on the evolution of the universe.

\section*{Acknowledgments}
This research was funded by the Science Committee of the Ministry of Science and Higher Education of the Republic of Kazakhstan (Grant No. AP22682760).

\section*{Data Availability Statement}
This article does not introduce any new data.

%%%%%%%%%%%%%%%%%%%%%%%%%%%%%%%%%%%%%%%%%%%%%%%%%%%%%%%%%%%%%%%%%%%%%%%%%%%%%%%%%
%%
%%


\begin{thebibliography}{90}

\bibitem{Riess/1998} A.G. Riess et al., Astron. J., \textbf{116}, 1009 (1998).

\bibitem{Perlmutter/1999} S. Perlmutter et al., Astrophys. J., \textbf{517}, 565 (1999). 

\bibitem{WMAP/2003} WMAP Collaboration, G. Hinshaw et al., ApJ. Suppl. Ser., \textbf{208}, 19 (2013). 

\bibitem{Planck/2014} Planck Collaboration XVI, Ade P.A.R. et al., Astron. Astrophys., \textbf{571}, A16 (2014). 

\bibitem{Planck/2015} Planck Collaboration XIII, Ade P.A.R. et al., Astron. Astrophys., \textbf{594}, A13 (2015).

\bibitem{Planck/2016} Planck Collaboration XIV, Ade P.A.R. et al., Astron. Astrophys., \textbf{594}, A14 (2016).

\bibitem{Planck/2020} Planck Collaboration VI, N. Aghanim et al., Astron. Astrophys., \textbf{641}, A6 (2020).

\bibitem{Weinberg/1989} S. Weinberg, Rev. Mod. Phys., \textbf{61}, 1 (1989).

\bibitem{Sahni/2000} V. Sahni and A.A. Starobinsky, Int. J. Mod. Phys. A, \textbf{9}, 373 (2000).

\bibitem{Padmanabhan/2003} T. Padmanabhan, Phys. Rept., \textbf{380}, 235 (2003).

\bibitem{Peebles/2003} P.J.E. Peebles and B. Ratra, Rev. Mod. Phys., \textbf{65}, 559 (2003).

\bibitem{Copeland/2006} E.J. Copeland, M. Sami, and S. Tsujikawa, Int. J. Mod. Phys. D, \textbf{15}, 1753 (2006).

\bibitem{Einstein/1917} A. Einstein, Kosmologische betrachtungen zur allgemeinen relativit\"atstheorie, Sitzungsber. Kgl. Preuss. Akad. Wiss., 142-152 (1917).

\bibitem{Amendola/2000} L. Amendola, Phys. Rev. D, \textbf{62}, 043511 (2000).

\bibitem{Maia/2002} J.M.F. Maia and J.A.S. Lima, Phys. Rev. D, \textbf{65}, 083513 (2002).

\bibitem{Koivisto/2005} T. Koivisto, Phys. Rev. D, \textbf{72}, 043516 (2005).

\bibitem{Lee/2006} S. Lee, G. C. Liu and K. W. Ng, Phys. Rev. D, \textbf{73}, 083516 (2006).

\bibitem{Bertolami/2007} O. Bertolami, F. Gil Pedro and M. Le Delliou, Phys. Lett. B, \textbf{654}, 165 (2007).

\bibitem{Ozer/1986} M. \"Ozer and M.O. Taha, Phys. Lett. B, \textbf{171}, 363 (1986).

\bibitem{Freese/1987} K. Freese et al., Nucl. Phys. B, \textbf{287}, 797 (1987).

\bibitem{Chen/1990} W. Chen and Y-S. Wu, Phys. Rev. D, \textbf{41}, 695 (1990).

\bibitem{Berman/1991} M.S. Berman, Phys. Rev. D, \textbf{43}, 1075 (1991).

\bibitem{Pavon/1991} D. Pav\'on, Phys. Rev. D, \textbf{43}, 375 (1991).

\bibitem{Carvalho/1992} J.C. Carvalho, J.A.S. Lima and I. Waga, Phys. Rev. D, \textbf{46}, 2404 (1992).

\bibitem{Arbab/1994} A.I. Arbab and A.M.M. Abdel-Rahman, Phys. Rev. D, \textbf{50}, 7725 (1994).

\bibitem{Barrow/2006} J.D. Barrow and T. Clifton, Phys. Rev. D, \textbf{73}, 103520 (2006).

\bibitem{Wang/2006} B. Wang, C.Y. Lin, and E. Abdalla, Phys. Lett. B, \textbf{637}, 357 (2006).

\bibitem{Montenegro/2007} A.E. Montenegro Jr. and S. Carneiro, Class. Quant. Grav., \textbf{24}, 313 (2007).

\bibitem{Overduin} J.M. Overduin and F.I. Cooperstock, Phys. Rev. D, \textbf{58}, 043506 (1998).

\bibitem{Rezaei} M. Rezaei, M. Malekjani, and J.S. Peracaula, Phys. Rev. D, \textbf{100}, 023539 (2019).

\bibitem{Carneiro/2003} S. Carneiro, Int. J. Mod. Phys. D, \textbf{12}, 1669 (2003).

\bibitem{Alcaniz/2005} J.S. Alcaniz and J.A.S. Lima, Phys. Rev. D, \textbf{72}, 063516 (2005).

\bibitem{Sola/2008} J. Sol\`a, J. Phys. A, \textbf{41}, 164066 (2008).

\bibitem{Sola/2011} J. Sol\`a, J. Phys. Conf. Ser., \textbf{283}, 012033 (2011).

\bibitem{Sola/2013} J. Sol\`a,  J. Phys. Conf. Ser., \textbf{453}, 012015 (2013).

\bibitem{Carneiro/2008} S. Carneiro, Phys. Rev. D, \textbf{77}, 083504 (2008).

\bibitem{Ryden} B. Ryden, Introduction to Cosmology; Addison Wesley: San Francisco, \textbf{CA}, USA (2003).

\bibitem{Koussour1} M. Koussour et al., Results Phys., \textbf{55}, 107166 (2023).

\bibitem{Myrzakulov1} N. Myrzakulov, M. Koussour and D.J. Gogoi, Eur. Phys. J. C, \textbf{83}, 594 (2023).

\bibitem{Wang/2017} D. Wang and X.H. Meng,  Phys. Rev. D, \textbf{96}, 103516 (2017).

\bibitem{Mackey/2013} D. F. Mackey et al., \textit{Publ. Astron. Soc. Pac.} \textbf{125}, 306 (2013).

\bibitem{baye}M.P. Hobson, A.H. Jaffe, A.R. Liddle, P. Mukherjee, D. Parkison (Eds.), Bayesian Methods in Cosmology, Cambridge University Press, Cambridge, England (2009).

\bibitem{Jimenez:2003iv} R. Jimenez et al., Astrophys. J., \textbf{593}, 622-629 (2003).

\bibitem{Simon:2004tf} J. Simon, L. Verde, and R. Jimenez, Phys. Rev. D, \textbf{71}, 123001 (2005).

\bibitem{Stern:2009ep} D. Stern et al., JCAP, \textbf{02}, 008 (2010).

\bibitem{Moresco:2012jh} M. Moresco et al., JCAP, \textbf{08}, 006 (2012).

\bibitem{Zhang:2012mp} C. Zhang et al., Res. Astron. Astrophys., \textbf{14}, 1221-1233 (2014).

\bibitem{Moresco:2015cya} M. Moresco, Mon. Not. Roy. Astron. Soc., \textbf{450}, L16-L20 (2015).

\bibitem{Moresco:2016mzx} M. Moresco et al., JCAP, \textbf{05}, 014 (2016).

\bibitem{Ratsimbazafy:2017vga} A.L. Ratsimbazafy et al., Mon. Not. Roy. Astron. Soc., \textbf{467}, 3239-3254 (2017).

\bibitem{pantheon}D. M. Scolnic, et al., The Complete Light-curve Sample of Spectroscopically Confirmed SNe Ia from Pan-STARRS1 and Cosmological Constraints from the Combined Pantheon Sample, Astrophys. J. \textbf{859}, 101. (2018).

\bibitem{Riess/2019} A.G. Riess et al., Astrophys. J., \textbf{876}, 85 (2019).

\bibitem{Valentino/2021A} E.D. Valentino et al., Astropart. Phys., \textbf{131}, 102605 (2021).

\bibitem{Yang/2021} W. Yang et al., Phys. Rev. D, \textbf{104}, 063521 (2021).

\bibitem{Valentino/2021B} E.D. Valentino, S. Pan, W. Yang, and L.A. Anchordoqui, Phys. Rev. D, \textbf{103}, 123527 (2021).

\bibitem{Koussour2} M. Koussour et al., Phys. Dark Universe, \textbf{42}, 101339 (2023).

\bibitem{Koussour3} M. Koussour et al., Eur. Phys. J. C, \textbf{83}, 1-14 (2023).

\bibitem{Koussour4} N. Myrzakulov et al., Eur. Phys. J. Plus, \textbf{138}, 852 (2023).

\bibitem{Koussour5} N. Myrzakulov et al., Chinese Phys. C, \textbf{47}, 115107 (2023).

\bibitem{Farooq/2013} O. Farooq and B. Ratra, Astrophys. J. Lett., \textbf{766}, L7 (2013).

\bibitem{Lu/2011} J. Lu, L. Xu, and M. Liu, Phys. Lett. B, \textbf{699}, 246 (2011).

\bibitem{Yang/2020} Y. Yang and Y. Gong, J. Cosmol. Astropart. Phys., \textbf{06}, 059 (2020).

\bibitem{Capozziello/2014} S. Capozziello, O. Farooq, O. Luongo, and B. Ratra, Phys. Rev. D, \textbf{90}, 044016 (2014).

\bibitem{Capozziello/2020} S. Capozziello, R. D'Agostino, and O. Luongo, Mon. Not. R. Astron. Soc., \textbf{494}, 2576 (2020).

\bibitem{Gruber/2014} C. Gruber and O. Luongo, Phys. Rev. D, \textbf{89}, 103506 (2014).

\bibitem{Sahni/2008} V. Sahni, A. Shafieloo and A.A. Starobinsky, Phys. Rev. D, \textbf{78}, 103502 (2008).

\bibitem{Sahni/2014} V. Sahni, A. Shafieloo, and A.A. Starobinsky, Astrophys. J., \textbf{793}, L40 (2014).

\bibitem{Shahalam/2015} M. Shahalam, S. Sami, and A. Agarwal, Mon. Not. R. Astron. Soc., \textbf{448}, 2948 (2015).

\bibitem{Sahni/2003} V. Sahni, T.D. Saini, A.A. Starobinsky, and U. Alam, JETP Lett., \textbf{77}, 201 (2003).

\bibitem{Alam/2003} U. Alam, V. Sahni, T.D. Saini, and A.A. Starobinsky, Mon. Not. R. Astron. Soc., \textbf{344}, 1057 (2003).

\bibitem{Mukherjee/2016} A. Mukherjee and N. Banerjee, Phys. Rev. D, \textbf{93}, 043002 (2016).

\bibitem{Zhai/2013} Z. X. Zhai et al., Phys. Lett. B, \textbf{727}, 8 (2013).

\bibitem{Akarsu/2014} \"O. Akarsu, S. Kumar, R. Ayrzakulov, M. Sami, and L. Xu, J. Cosmol. Astropart. Phys., \textbf{1}, 022 (2014).

\bibitem{Mamon/2018} A.A. Mamon and K. Bamba, Eur. Phys. J. C, \textbf{78}, 862 (2018).



\end{thebibliography}
\end{document}